\documentclass[3p,times,procedia]{elsarticle}
\usepackage{nupha_ecrc}

\volume{00}

\firstpage{1}

\journalname{Nuclear Physics A}

\runauth{B. Schenke}

\jid{nupha}

\jnltitlelogo{Nuclear Physics A}

\usepackage{amssymb}

\usepackage[figuresright]{rotating}
\usepackage[utf8]{inputenc}

\begin{document}

\begin{frontmatter}



\dochead{XXVIth International Conference on Ultrarelativistic Nucleus-Nucleus Collisions\\ (Quark Matter 2017)}

\title{Origins of collectivity in small systems}

\author{Bj\"orn Schenke}
\address{Physics Department, Brookhaven National Laboratory, Upton, NY 11973, USA}
\ead{bschenke@bnl.gov}


\begin{abstract}
We review recent developments in the theoretical description and understanding of multi-particle correlation measurements in collisions of small projectiles (p/d/$^3$He) with heavy nuclei (Au, Pb) as well as proton+proton collisions. We focus on whether the physical processes responsible for the observed long range rapidity correlations and their azimuthal structure are the same in small systems as in heavy ion collisions. In the latter they are interpreted as generated by the initial spatial geometry being transformed into momentum correlations by strong final state interactions. However, explicit calculations show that also initial state momentum correlations are present and could contribute to observables in small systems. If strong final state interactions are present in small systems, recent developments show that results are sensitive to the shape of the proton and its fluctuations.
\end{abstract}


\end{frontmatter}


\section{Introduction}
Multi-particle correlation measurements in collisions of small projectiles (e.g. p/d/$^3$He) with other small projectiles or heavy ions (e.g. Au, Pb) show characteristic structures that are long range in rapidity and have azimuthal anisotropies very similar to what was measured in heavy ion collisions 
\cite{Dusling:2015gta}. In heavy ion collisions these structures have for a long time been interpreted as emerging from the system's response (via strong final state interactions) to the initial shape of the interaction region \cite{Heinz:2013th,Gale:2013da,deSouza:2015ena,Song:2017wtw}. In this case the long range nature of the correlation is due to the transverse geometry being almost rapidity independent. 
To model the final state either relativistic hydrodynamics or transport simulations have been employed. When combined with sophisticated event-by-event initial state models, very good agreement with a wide range of experimental data, including the correlations under consideration, has been achieved with these calculations \cite{Gale:2013da}.

The natural question to ask is whether multi-particle correlation measurements in small collision systems are described within the same framework and thus can be interpreted to have the same physical origin. The reason why the answer to that question is not a simple yes is two-fold. First, the applicability of some of the most successful final state simulations, namely relativistic viscous hydrodynamics, becomes more questionable as the system size decreases and gradients become larger. Second, a wide range of calculations, mostly within the color glass condensate (CGC) framework, which is an effective theory of quantum chromo dynamics (QCD) at high energy, show that initial state momentum correlations of produced gluons are present and have similar structures as the experimentally observed correlations \cite{Dusling:2015gta,Schlichting:2016kjw}.

In the following we discuss the question of applicability of hydrodynamics, the role of subnucleonic fluctuations, and the contribution from initial state momentum correlations. We then follow with a discussion of what observables could give further insight into what source of correlation dominates in a given system for a given multiplicity range.

\section{Applicability of hydrodynamics and the role of non-equilibrium evolution}
Even in the case of heavy ion collisions applying hydrodynamics to describe the system at relatively early times has been under debate. This is because to this date it has not been understood from first principles how the system can evolve from very far from equilibrium (two colliding nuclei moving in opposite directions) to close to locally equilibrated, or at least close to locally isotropic \cite{Arnold:2004ti} quickly enough. In fact, in a rapidly expanding system like heavy ion collisions, no theory, neither in the strong nor weak coupling limit, has ever found pressure anisotropies (between longitudinal and transverse pressures) that are smaller than 50\% \cite{Romatschke:2016hle}.

Yet, for reasons unknown, hydrodynamics describes the system even for large pressure anisotropies extremely well \cite{Chesler:2009cy,Heller:2011ju,Casalderrey-Solana:2013aba,Kurkela:2015qoa,Keegan:2015avk}. A recent attempt \cite{Romatschke:2015gic,Romatschke:2016hle} at an explanation for this behavior is based on studies of the hydrodynamic and non-hydrodynamic modes of various theories (e.g. in AdS/CFT or kinetic theory) \cite{Heller:2013fn,Heller:2015dha,Buchel:2016cbj,Denicol:2016bjh,Heller:2016rtz}. It is argued that the system behaves hydrodynamically as long as the hydrodynamic modes (for which $\omega(k) \rightarrow 0$ as $k\rightarrow 0$) dominate, independent of the system's isotropy. Within this interpretation a lower limit of $\sim 0.15\,{\rm fm}$ for the size of a droplet of fluid nuclear matter was extracted \cite{Romatschke:2016hle}.

Even if hydrodynamics is applied at times $\gtrsim 0.5\,{\rm fm}$ the details of the non-equilibrium evolution before that time will be more important in small systems, whose total evolution time is often only $\lesssim 3\,{\rm fm}$. This increased importance of the non-equilibrium stage has been demonstrated for photon production in \cite{Berges:2017eom}. Thus, better theoretical understanding is needed of this hard to describe part of the evolution. Recent progress in this direction has been made in both the weak and strong coupling limits. See \cite{Kurkela:2016vts} and \cite{Mrowczynski:2016etf} for recent reviews.

Final state interactions can also be simulated microscopically without relying on the applicability of hydrodynamics. Calculations performed in the AMPT framework, which uses a parton cascade to describe the final state interactions, can also reproduce the anisotropic momentum distributions of produced particles in small systems \cite{Ma:2014pva,Bzdak:2014dia,Koop:2015wea,Bozek:2015swa,He:2015hfa}. Interestingly the data is well described when using a rather small parton-parton cross section. This leads to an interpretation dubbed ``parton escape mechanism'', which is fundamentally different from the collective motion produced in hydrodynamics. In the AMPT simulations only a few collisions are required to produce the momentum anisotropy, which emerges because partons are more likely not to scatter, and thus escape, if the medium is shorter in the direction they are moving. It is not clear at this point how to distinguish the two scenarios experimentally.

\section{Subnucleonic fluctuations and their effect on anisotropic flow in p+Pb collisions}
Assuming that hydrodynamics is in fact applicable to describe the final state dynamics of small collision systems, many different calculations have produced results for anisotropic flow coefficients comparable to the experimental data from the LHC (see the review \cite{Dusling:2015gta} and more recent work \cite{Shen:2016zpp,Weller:2017tsr}). The available calculations differ mainly in the choice of initial state, which in proton+proton and proton+heavy ion collisions is not well constrained. Calculations that employ variations of the Monte-Carlo Glauber model produced results compatible with most of the data, even when subnucleonic fluctuations of the proton were ignored. However, the IP-Glasma model \cite{Schenke:2012wb,Schenke:2012hg} for example, which is one of the few initial state models not excluded by data from heavy ion collisions, such as the event-by-event distributions of anisotropic flow coefficients, could not reproduce the momentum anisotropy in p+Pb collisions at the LHC \cite{Schenke:2014zha}. The reason for the disagreement was the assumption of an approximately (modulo small color charge fluctuations) round proton, whose shape dominates the shape of the interaction region in the IP-Glasma's Yang-Mills framework.

Recent focus has been on understanding possible sub-nucleonic structure fluctuations and within the IP-Glasma model it was found that also describing data on incoherent diffractive vector meson production in electron+proton collisions at HERA requires additional fluctuations of the proton structure \cite{Mantysaari:2016ykx,Mantysaari:2016jaz}. Introducing hot spots in the gluon distribution, e.g. by assuming that gluons are lumped around positions of three valence quarks, leads to increased fluctuations of the scattering amplitude and thus the incoherent cross section. Then parameters, such as the size of the proton and the hot-spots can be tuned to describe the HERA data, just as HERA data for e.g. $F_2$ was used to constrain the original IP-Glasma model. It was found that when these fluctuating protons are used in the calculation of p+Pb collisions, anisotropic flow coefficients increase by up to a factor of 5, being compatible with the experimental data \cite{Mantysaari:2017wtb,Mantysaari:2017cni}. This is due to the much increased shape fluctuations of the interaction region, necessary to generate large anisotropic flow when an average anisotropy of the geometry is absent.

Further interesting developments towards understanding the fluctuating subnucleonic structure and its effect on observables in proton+proton and proton+heavy-ion collisions are underway. Recent work has focused on including wounded quarks instead of nucleons in Glauber type models \cite{Glazek:2016vkl,Bozek:2016kpf,Welsh:2016siu,Weller:2017tsr,Bozek:2017elk}, proton size fluctuations \cite{McGlinchey:2016ssj}, and shape fluctuations driven by fluctuations of the proton spin \cite{Habich:2015rtj}. Also Bayesian analyses \cite{Bernhard:2016tnd} are being extended to include subnucleonic structures \cite{Moreland:2017kdx}.

\section{Initial state momentum correlations}
Most calculations concerned with momentum correlations that are intrinsic to the initially produced particle distributions are based on the color glass condensate (CGC) effective theory of QCD \cite{Gelis:2010nm}. Just as in the IP-Glasma framework mentioned above, the production of gluons in nuclear collisions is determined from solutions of the Yang-Mills equations given the incoming color currents of the two nuclei. The wide range of calculations of initial momentum correlations only differ in the approximations made to solve these equations. The first calculations were done in the so called ``glasma graph approximation'', which does not include more than two-gluon exchanges and uses Gaussian statistics for the initial color charges \cite{Gelis:2008ad,Gelis:2008sz,Dumitru:2008wn,Dumitru:2010mv,Dusling:2012wy,Dusling:2013qoz}.
The non-linear Gaussian approximation also uses Gaussian statistics but resums multi-gluon exchanges \cite{McLerran:1998nk,Dominguez:2008aa,Lappi:2015vta}, while fully numerical calculations \cite{Krasnitz:1998ns,Krasnitz:2002mn,Lappi:2003bi,Schenke:2012wb} include multi-gluon exchanges and allow for any color charge statistics and realistic spatial distributions \cite{Schenke:2015aqa}. Finally, one can go beyond the classical approximation by performing small-$x$ evolution such as JIMWLK \cite{Jalilian-Marian:1997jx,Jalilian-Marian:1997gr,Iancu:2000hn,Mueller:2001uk} on the initial color charge distributions \cite{Lappi:2015vha}.

All of these calculations have in common that they find initial anisotropy coefficients $v_n$ for even $n$. Odd $n$ are generated by gluon ``rescattering'' which is part of e.g. the numerical solution of the Yang-Mills equations when including the evolution in time \cite{Schenke:2015aqa}.

The color domain model \cite{Kovner:2010xk,Kovner:2011pe,Dumitru:2014dra,Dumitru:2014yza} amounts to introducing additional non-Gaussian contributions \cite{Lappi:2015vta}, which can arise from going beyond the glasma graph approximation, where $E^a\propto\rho^a$ (such that the color electric field $E$ has Gaussian correlators if the color charges $\rho$ do), the JIMWLK evolution of a Gaussian initial condition, or intrinsic correlations in the initial condition at large $x$. The latter could emerge from e.g. a finite number of sources \cite{McLerran:2015sva}.

In the following, we discuss which experimental observables can help to determine the physical process responsible for the observed correlations in small systems. Certain characteristic features of the initial momentum correlations and the final state dominated correlations, like the (in-)dependence on the global initial geometry will play an important role in this process.

\section{Observables to distinguish the two scenarios}
Many observables have been suggested and dedicated experiments conducted to distinguish whether the correlations observed in small systems originate from initial spatial correlations, which are then converted to final state momentum correlations, or from intrinsic initial state momentum correlations. Most observables were chosen because of the specific expectations one would have for them in the hydrodynamic picture. This includes the particle mass dependence of the mean transverse momentum and anisotropy coefficients, the equality of higher order cumulants, and the correlation between the initial geometry and the observed anisotropies. In the following we will list the most revealing observables and discuss the current understanding they provide.

\emph{Mass ordering} The characteristic particle mass dependence of the mean transverse momentum has been discussed as a signature of a strongly interacting medium for a long time \cite{Levai:1991be}. However, other explanations not relying on hydrodynamic behavior were also given early on \cite{Wang:1991vx}.
The situation is similar for the mass ordering of anisotropy coefficients $v_n$. It has been shown that for small systems, the experimentally observed mass splitting is well reproduced by hydrodynamic simulations \cite{Bozek:2013ska,Werner:2013ipa}. However, recent calculations combining the initial state gluon momentum anisotropy from the IP-Glasma with a Lund string fragmentation framework, also reproduce the mean transverse momentum and $v_2$ mass splitting observed in p+p collisions at the LHC \cite{Schenke:2016lrs}. In both scenarios particles are emitted from a common source moving with a given velocity (fluid cells or moving strings, respectively), such that the observed mass dependent effects on the particle momentum spectra are very similar. So at this point the observation of mass ordering does not provide conclusive evidence for either scenario.

\emph{Sign change of the four-particle cumulant} A real valued $v_2\{4\}=\sqrt[4]{-c_2\{4\}}$ \cite{Bilandzic:2010jr} is only obtained for a negative four particle cumulant $c_2\{4\}$. In the hydrodynamic framework one has $c_2\{4\}=\langle\cos[n(\phi_1+\phi_2-\phi_3-\phi_4)]\rangle-2\langle\cos[n(\phi_1-\phi_2)]\rangle^2 = v_n^4-2(v_n^2)^2+ \dots= - v_n^4+\dots$ where the $\dots$ represent terms that appear due to fluctuations. The latter can change the sign of $c_2\{4\}$, such that even the hydrodynamic picture can produce positive $c_2\{4\}$. 
The experimental data of $c_2\{4\}$ as a function of multiplicity (in high energy p+p \cite{Khachatryan:2016txc} and p+Pb \cite{Chatrchyan:2013nka} collisions) shows a clear change in sign, which is often interpreted as the indication of an onset of hydrodynamic behavior at a certain threshold in multiplicity. There are no calculations complete enough to describe the entire multiplicity dependence of $c_2\{4\}$ in either framework. In the glasma graph approximation $c_2\{4\}$ is always positive, but it has been argued that the addition of non-linear and non-Gaussian effects leads to a negative contribution to the four particle cumulant \cite{Dumitru:2014yza}. However, the exact multiplicity dependence of that contribution is not known and a quantitative description of the data has not yet been achieved.

\emph{Multi-particle cumulants} Hydrodynamics predicts the approximate equality of all higher order cumulants $c_n\{m\}$, with $m \geq 4$, because (for a given $n$) all particles are correlated with one common event plane (see e.g. \cite{Yan:2013laa}). Initial momentum space correlations do not obviously have the same feature. It was shown that in the dilute-dense limit of the color glass condensate framework and for transverse momenta much greater than $Q_s$, it holds that $|v_2\{2m\}|\approx |v_2\{2m'\}|$ for $m,m'\geq 4$ \cite{Skokov:2014tka}. However, harmonics with $2m=4n$ are imaginary, which is again potentially cured by including non-Gaussian correlations as discussed above for $m=4$.

\emph{Odd harmonics} Odd harmonics are generated in the hydrodynamic framework when fluctuations of the initial geometry are present. Their magnitudes depend on the initial state model used. It can be expected that they depend on the proton substructure in p+p and p+A collisions \cite{Schenke:2014zha}.
We mentioned above that within the usually used approximations odd harmonics for gluons from the initial state are generated only via final state interactions, e.g. via evolution of the Yang-Mills fields \cite{Schenke:2015aqa}. This has also been demonstrated analytically in \cite{McLerran:2016snu}. When going beyond the classical and dilute limits simultaneously, recent work has demonstrated that odd harmonics can also be present in the initial state momentum distributions \cite{Gotsman:2016fee,Gotsman:2016wtq1,Kovner:2016jfp}. 

\begin{figure}
\begin{minipage}[t]{0.48\textwidth}
\begin{center}
  \includegraphics[width=\textwidth]{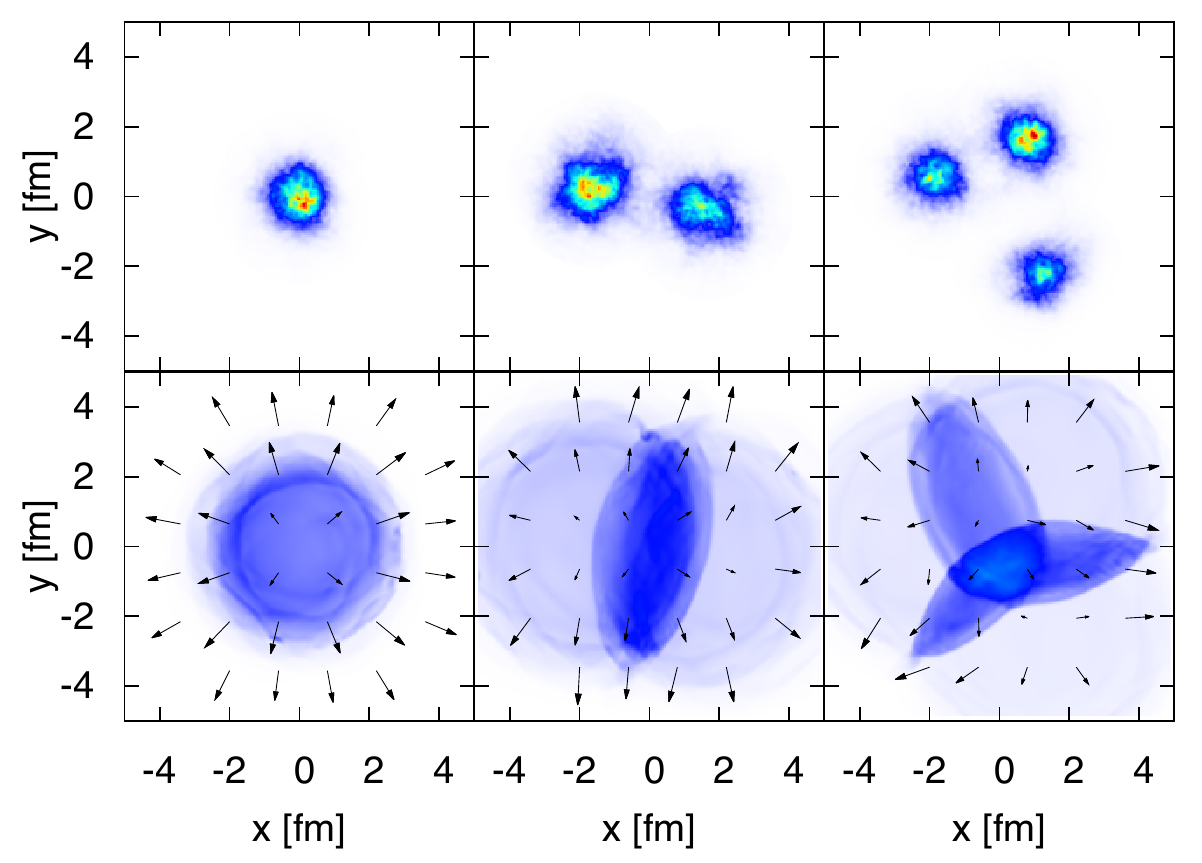} 
  \caption{ Upper panel: IP-Glasma initial configurations of the energy density in single events of p+Au, d+Au and $^3$He+Au collisions, respectively. Lower panel: Energy density and flow directions (arrows) after hydrodynamic evolution of the IP-Glasma initial conditions above. Characteristic elliptic and triangular flow patterns are visible in the middle and right panel, respectively. \label{fig:thefigure}}
\end{center}
\end{minipage}
\hspace{0.04\textwidth}
\begin{minipage}[t]{0.48\textwidth}
\begin{center}
 \includegraphics[width=\textwidth]{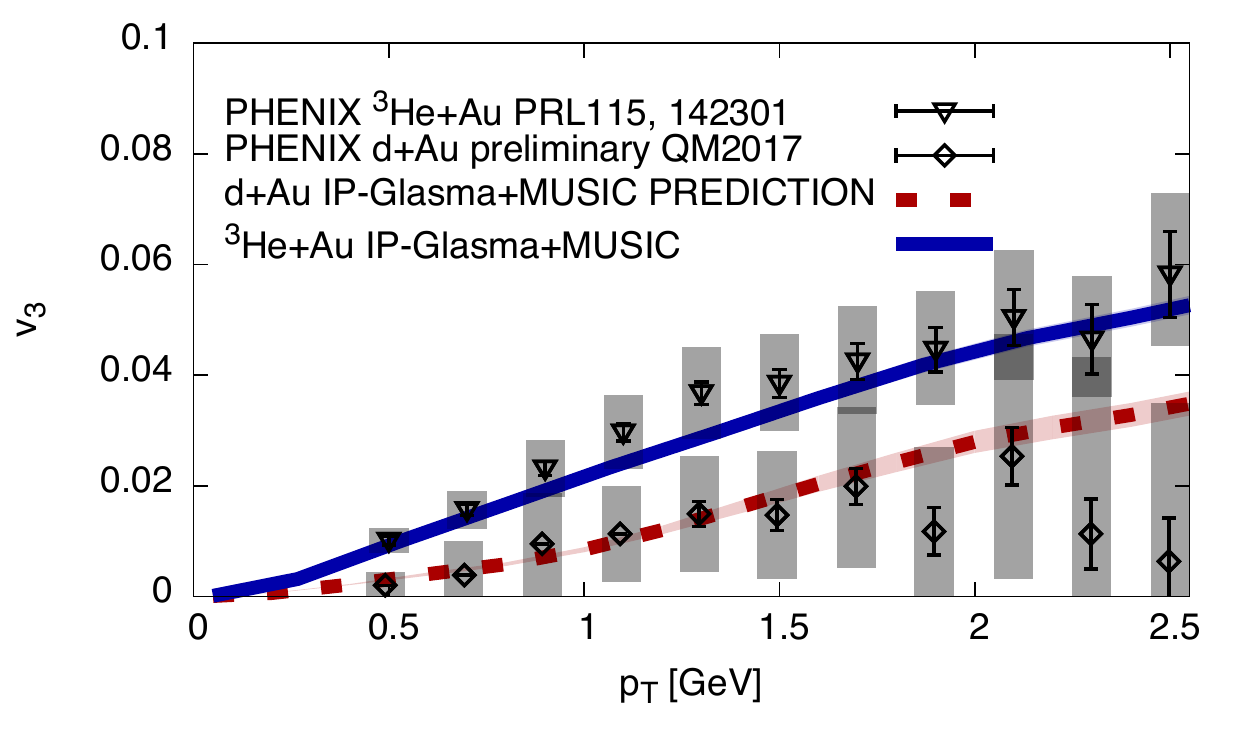}
  \caption{ Triangular flow coefficients $v_3$ in d+Au and $^3$He+Au collisions from the IP-Glasma+\textsc{Music} framework compared to experimental data from the PHENIX Collaboration \cite{Adare:2015ctn,McGlinchey:2017esf}. \label{fig:thevns}}
\end{center}
\end{minipage}
\end{figure}

\emph{Different collision systems} In the scenario where final state interactions dominate the observed momentum space correlations, one is very sensitive to the initial shape of the collision system. If one modifies the average geometry in small systems by using varying small projectiles, like proton, deuteron, or $^3$He, one expects a modification of the measured ellipticity and triangularity coefficients for example (see Fig. \ref{fig:thefigure} for example events in three different collision systems). RHIC has performed a study of these coefficients in different small systems and comparison with hydrodynamic calculations confirmed the expectations \cite{Adare:2014keg,Adare:2015ctn,Romatschke:2015gxa}. 
Within the IP-Glasma+\textsc{Music} framework predictions were made for all three different collision systems \cite{Schenke:2014gaa}. After comparing to new $^3$He data for $v_3$ from PHENIX \cite{Adare:2015ctn}, transport parameters were adjusted and predictions for $v_3$ in d+Au collisions made. The results are compared to new preliminary data from PHENIX \cite{McGlinchey:2017esf} in Fig.\,\ref{fig:thevns}. Agreement between the predicted $v_3$ in d+Au collisions, which is significantly smaller than that in $^3$He+Au collisions, and experimental data is very good. 

Whether the initial state framework can describe such trends remains to be seen. If it is the case then the origin of the change in $v_3$ for example must be different than in the final state framework. This is because it has been shown \cite{Schenke:2015aqa,McLerran:2016snu} that there is no correlation between the global event geometry and the produced momentum anisotropy in the initial state picture.

\emph{Jets and electromagnetic probes} One expects jet quenching like in heavy ion collisions if a strongly interacting medium is formed in small collision systems. Some recent calculations have combined jet energy loss Monte Carlo with a hydrodynamic background describing the bulk of p+Pb collisions \cite{Shen:2016egw}. It was found for minimum bias events in p+Pb collisions that the nuclear modification factor $R_{\rm pA}$ is consistent with one, even when jet energy loss is included. Significant suppression is predicted for 0-1\% central events, but comparison to experimental data is not possible before an unambiguous centrality selection can be performed experimentally \cite{ATLAS:2014cpa}. Also energy loss in cold nuclear matter should be considered as a possibility in small systems \cite{Kang:2015mta}.

If a (close to) thermal medium is produced in small collision systems one expects additional contributions to direct photon production as in heavy ion collisions. These thermal photons have been shown to lead to an up to a factor of two enhancement of the direct photon yield in 0-5\% central p+Pb collisions, depending on the transverse photon momentum \cite{Shen:2016zpp}. Uncertainties in the extraction of direct photon yields are large, so it remains to be seen if this enhancement can be measured. Also, as mentioned above, non-equilibrium photon production may play an important role \cite{Berges:2017eom}.

\section{Conclusions}
The detailed analysis of small collision systems has led to many new physics insights, including the nature of multi-particle production from QCD at high energy. There is no question that non-trivial momentum space correlations exist in the initial multi-gluon distribution. It remains to be seen in what collision systems and at what multiplicities these can survive possible strong final state interactions. It is expected that as the multiplicity decreases, initial state effects rise in prominence. Many qualitative (and to some degree quantitative) explanations that employ the final state framework, where correlations emerge from the conversion of initial state spatial structures, have been presented. This indicates that the studied high multiplicity events in small systems behave similarly to heavy ion collisions, potentially leading to new insight into the fluctuating spatial structure of hadrons and small nuclei.

\section*{Acknowledgments}
The author thanks Raju Venugopalan for helpful comments on the manuscript. BPS is supported by the U.S. Department of Energy under Grant No. DE-SC0012704.
Numerical calculations used resources of NERSC, a DOE Office of Science User Facility supported by the Office of Science of the U.S. Department of Energy under Contract No. DE-AC02-05CH11231.





\bibliographystyle{elsarticle-num}
\bibliography{spires}







\end{document}